\documentclass{elsart}
\usepackage[dvips]{graphicx}
\usepackage{amsmath}              
\usepackage{amssymb}              
\usepackage{amsbsy}              
\newcommand{\kstar}{\ensuremath{D^{+}\to \overline{K}^{*0}\mu^{+}\nu}}
\newcommand{\rhomunu}{\ensuremath{D^{+}\to \rho^{0}\mu^{+}\nu}}

\newcommand{\kpipi}{\ensuremath{D^{+}\to K^{-}\pi^{+}\pi^{+}}}

\newcommand{\dsetamunu}{\ensuremath{D^{+}_{s}\to\eta\mu\nu}}
\newcommand{\dsetaprimemunu}{\ensuremath{D^{+}_{s}\to\eta'\mu\nu}}

\newcommand{\phimunu}{\ensuremath{D^{+}_{s}\to\phi\mu^+\nu}}
\newcommand{\threebody}{\ensuremath{M(\pi^+\pi^-\mu^+)}}
\newcommand{\twobody}{\ensuremath{M(\pi^+\pi^-\mu^+) - M(\pi^-\mu^+)}}
\newcommand{\kstarbr}{\ensuremath{\frac{\textrm{BR}(\kstar)}{\textrm{BR}(\kpipi)}}}
\newcommand{\rhobr}{\ensuremath{\frac{\textrm{BR}(\rhomunu)}{\textrm{BR}(\kstar)}}}
\newcommand{\rholnu}{\ensuremath{D^{+}\to \rho^{0}\ell^+\nu}}
\newcommand{\kstarlnu}{\ensuremath{D^{+}\to \overline{K}^{*0}\ell^+\nu}}
\newcommand{\rhoenu}{\ensuremath{D^{+}\to \rho^{0}e^{+}\nu}}
\newcommand{\kstarenu}{\ensuremath{D^{+}\to \overline{K}^{*0}e^{+}\nu}}
\newcommand{\kpimunu}{\ensuremath{D^+\to K^-\pi^+\mu^+\nu}}
\newcommand{\cer}{\v{C}erenkov}

\newcounter{saveeqn}%

\begin{document}
\begin{frontmatter}
\title{\boldmath New Measurement of $\frac{\textrm{BR} \left(D^{+}\to \rho^{0}\mu^{+}\nu \right)}{\textrm{BR} \left( D^{+}\to %
       \overline{K}^{*0}\mu^{+}\nu \right) }$ ~Branching Ratio}

\collaboration{The~FOCUS~Collaboration}\footnotemark
\author[ucd]{J.~M.~Link}
\author[ucd]{P.~M.~Yager}
\author[cbpf]{J.~C.~Anjos}
\author[cbpf]{I.~Bediaga}
\author[cbpf]{C.~Castromonte}
\author[cbpf]{A.~A.~Machado}
\author[cbpf]{J.~Magnin}
\author[cbpf]{A.~Massafferri}
\author[cbpf]{J.~M.~de~Miranda}
\author[cbpf]{I.~M.~Pepe}
% ...

\address[ucd]{University of California, Davis, CA 95616}
\address[cbpf]{Centro Brasileiro de Pesquisas F\'\i sicas, Rio de Janeiro, RJ, Brazil}
% ...

\footnotetext{See \textrm{http://www-focus.fnal.gov/authors.html} for additional author information.}

\nobreak
\begin{abstract}
   Using data collected by the FOCUS experiment at Fermilab, we present a new
   measurement of the charm semileptonic branching ratio \rhobr. 
   From a sample of $320\pm44$ and $11372\pm161$ \rhomunu ~and \kpimunu 
   ~events respectively, we find
   $\rhobr=0.041\pm0.006~\textrm{(stat)}\pm0.004~\textrm{(syst)}$.   
\end{abstract}
\end{frontmatter}

\section{Introduction}

   Semileptonic decays provide an ideal environment for the study of hadronic matrix elements affecting 
   the weak mixing angles from heavy flavor decays since the weak part of the current can be separated 
   from the hadronic part of the current. The hadronic current, described by form factors, can
   be calculated by different theoretical methods, e.g., Lattice QCD (LQCD), Quark Model (QM), 
   Sum Rules (SR), without the added complication of significant final state interactions.
   While many theoretical models predict the value for the branching ratio 
   $\frac{\textrm{BR}(\rholnu)}{\textrm{BR}(\kstarlnu)}$ ~\cite{%
   ball,ape,jaus,isgw2,yang_hwang,odonnell_turan,melikhov,ligeti,kondratyuk,melikhov-stech,wang_wu_zhong,kamenik}, 
   only a few experimental measurements of 
   this ratio have been made. Furthermore, even though previous measurements of \rhobr ~have suffered 
   from a lack of  statistics \cite{e653_rho,e687_rho,e791_rho}, most theoretical predictions 
   still differ by at least $2\sigma$ from the world average~\cite{pdg} for the muonic mode. 
   In this paper, we present a new measurement of this branching ratio based on $320\pm44$ \rhomunu 
   ~events.\footnote{Unless stated otherwise, charge conjugation is implied.} 
   
\section{Event Selection}

   The data for this analysis were collected with the FOCUS experiment during the
   1996--97 fixed target run at Fermilab. The FOCUS experiment utilized an upgraded version
   of the  forward multi-particle spectrometer used by experiment E687 \cite{e687_det} 
   to study charmed particles produced by the interaction of high energy photons, with an 
   average energy of $\sim180$ GeV, with a 
   segmented BeO target. Precise vertex determination was made possible by two sets of silicon strips
   detectors consisting of two pairs of planes interleaved with the target segments~\cite{TS}
   and four sets of planes downstream of the target region, each with three views. Five
   sets of proportional wire chambers combined with two oppositely polarized analysis
   magnets completed the tracking and momentum measurement system.
   Charged hadron identification was provided with three threshold multi-cell \cer 
   ~counters capable of separating kaons from pions up to 60 GeV/$c$~\cite{Cer}. Muons were 
   identified in the ``inner" muon system, located at the end of the spectrometer, which consisted
   of six arrays of scintillation counters subtending approximately $\pm 45$ 
   mrad~\cite{interference}. 
   
   The \rhomunu ~events are selected by requiring two oppositely charged pions and a
   muon to form a good decay vertex with confidence level exceeding 5\%. Tracks not used in
   the decay vertex are used to form candidate production vertices. Of these candidates, 
   the vertex with the highest multiplicity is selected as the production vertex; ties are broken
   by selecting the most upstream vertex. This production vertex 
   is required to have a confidence level greater than 1\% and be inside the target
   material. 
      
   Muon tracks are required to have a minimum momentum of 10 GeV/$c$ and must have hits in 
   at least five of the six planes comprising the inner muon system. These hits must be 
   consistent with the muon track hypothesis with a confidence level exceeding 1\%. In order to reduce 
   contamination from in-flight decays of pions and kaons within the spectrometer, we require the
   muon tracks to have a confidence level greater than 1\% under the hypothesis that the trajectory
   is consistent through the two analysis magnets.
   
   The \cer ~algorithm used for particle identification returns the negative log-likelihood
   for a given track to be either an electron, pion, kaon, or proton. To identify the two pion 
   candidates we require that no other hypothesis is favored over the pion hypothesis by 
   more than five units of log-likelihood for each track. Furthermore, we require that the pion 
   hypothesis for the track with charge opposite to the muon be favored over the kaon 
   hypothesis by at least five units of log-likelihood. This very stringent is cut used
   to suppress background from the Cabibbo favored decay $\kpimunu$ where
   the kaon is misidentified as a pion. Additionally, the pion hypothesis must
   be favored over the kaon hypothesis by more than one unit of log-likelihood for the remaining
   track.
   
   To suppress short-lived backgrounds, we require the decay vertex to be separated from the 
   production vertex by at least 15 times the calculated uncertainty on the separation $\sigma_L$
   and be outside of the target material by $1\sigma$. Because the target region has embedded detectors, 
   the decay vertex is also required to be outside the detector material. These cuts are
   especially effective at removing non-charm backgrounds, making the contribution from
   minimum-bias events negligible. Contamination from 
   higher multiplicity charm events is reduced by isolating the $\pi\pi\mu$ vertex from other
   tracks in the event (not including tracks in the primary). We require that the maximum
   confidence level for any other track to form a vertex with the candidate to be less 
   than 1\%.
   
   Background from $D^{*+} \to D^0\pi^+\to(\pi^-\mu^+\nu)\pi^+$, where the soft pion is 
   erroneously assigned to the decay vertex, is reduced by requiring 
   $\twobody >0.20~\textrm{GeV}/c^2$. Background decay modes with a neutral hadron in the final
   state, such as $D^+_s\to\eta'\mu^+\nu\to(\eta\pi^+\pi^+)\mu^+\nu$, are reduced by requiring 
   a visible mass cut
   of $1.2~\textrm{GeV}/c^2<\threebody< 1.8~\textrm{GeV}/c^2$. This cut eliminates 
   $\sim 40$\% of the \rhomunu ~signal, but also proves to be very effective at rejecting 
   background consisting mostly of kaons and pions misidentified as muons. 
      
   In order to reduce systematic errors common to both modes, the $\kpimunu$ events used for
   normalization had the same vertex and muon identification cuts as in the $\rhomunu$ analysis. 
   The kaon is identified by requiring that the kaon hypothesis is favored over the pion hypothesis
   by at least two units of log-likelihood. The pion candidate is identified by requiring 
   that the pion hypothesis for this track be favored over the kaon hypothesis.
   Background from $D^{*+} \to D^0\pi^+\to(K^-\mu^+\nu)\pi^+$ is reduced by requiring 
   $M(K^-\pi^+\mu^+) - M(K^-\mu^+) >0.20~\textrm{GeV}/c^2$. A cut on the visible mass of
   $1.0~\textrm{GeV}/c^2<M(K^-\pi^+\mu^+)<1.8~\textrm{GeV}/c^2$ suppresses background from muon
   misidentification. 
    
\section{BR Determination}  
   The \rhomunu ~yield is estimated using a binned maximum log-likelihood fit
   of the $\pi^+\pi^-$ invariant mass. The likelihood is defined as:
   \begin{equation}
      \mathcal{L}=\prod_{i=1}^{\textrm{\#bins}}
      \frac{n^{s_{i}}_{i}e^{-n_{i}}}{s_{i}!}\times\textrm{penalty}	
   \end{equation}							     	
   where $s_{i}$ is the number of events in bin ${i}$ of the data histogram,	
   $n_{i}$ is the number of events in bin ${i}$ of the fit histogram, and a penalty term, described
   below, is used to set a loose constraint on a known branching ratio.
   
   The fit histogram is composed of: binned, normalized shapes of signal and background components
   obtained from Monte Carlo simulations,
   $S_x$, the number of events as estimated by the fit for each shape, $Y_x$, and the number of events
   that occur due to feed-down from the Cabibbo-favored decay \kpimunu. The number of events in
   each bin is then:
   
   \begin{equation}
   \begin{split}
       n_{i} = &Y_{D^+\to\rho^0\mu^+\nu} S_{\rho^0\mu^+\nu}~+~  					
       \textrm{ECY}_{\kpimunu}\epsilon(K\pi\mu\nu\to\rho\mu\nu)S_{K\pi\mu\nu}~+\\	
      &Y_{D^+\to K^0_S\mu^+\nu} S_{K^0_S\mu^+\nu}~+~Y_{D^+\to\omega\mu^+\nu}
       S_{\omega\mu^+\nu}~+~Y_{D^+_s} S_{D^+_s}~+~
       Y_{\mathcal{C}}S_{\mathcal{C}}~+~Y_{\mathcal{M}}S_{\mathcal{M}}
   \end{split}
   \label{mll}
   \end{equation}
   where the terms in Eq.~\ref{mll} are explained in detail below.
   
   As mentioned before, the shapes for the signal and background are obtained via Monte Carlo
   simulation. The Monte Carlo is based on Pythia 6.127~\cite{pythia} and contains all known charm
   decays with their corresponding branching ratios and careful simulation of known secondary processes. 
   After an event is generated, it is passed through a simulation of the FOCUS spectrometer. The
   events are then selected in the same way as the data.
   
   $Y_{D^+\to\rho^0\mu^+\nu}$ is the yield of the \rhomunu ~signal. $\textrm{ECY}_{\kpimunu}$ is the
   efficiency-corrected yield (ECY) for \kpimunu. This quantity is the estimated number of \kpimunu
   ~events produced by FOCUS. This, along with the Monte  Carlo efficiency for a \kpimunu ~event to be 
   misidentified as a \rhomunu ~event, $\epsilon(K\pi\mu\nu\to\rho\mu\nu)$, provide an estimate of 
   the amount of feed-down of this mode into our signal. The ECY is fixed in the fit to the value 
   obtained from the $\kpimunu$ analysis used for the normalization mode. $Y_{D^+\to K^0_S\mu^+\nu}$ 
   is the yield of a small $K^0_S\to\pi^+\pi^-$ component. 
   
   $Y_{D^+\to\omega\mu^+\nu}$ is the yield of 
   $D^+\to\omega\mu^+\nu$, where the $\omega$ could decay either to $\pi^+\pi^-\pi^0$ or to
   $\pi^+\pi^-$. We use the recent CLEO--c collaboration measurements of the absolute branching ratio 
   of $D^+\to\overline{K}^{*0}e^+\nu$ and $D^+\to\omega e^+\nu$~\cite{cleo_d+} ~to set
   a loose constraint on the yield of $D^+\to\omega\mu^+\nu$. To this end, we add
   a penalty term to the likelihood of the form\footnote{Here we have assumed that the
   electronic and muonic rates are equal.}
   \begin{equation}
   \label{penalty}
   \exp{\biggl[-\frac{1}{2}\biggl( R_{\omega/\overline{K}^{*0}}
   \textrm{ECY}_{\kstar}~-~Y_{D^+\to\omega\mu^+\nu}\biggr)^2/%
   \sigma_{D^+\to\omega e\nu}^2}\biggr]
   \end{equation}
   
   where $R_{\omega/\overline{K}^{*0}} =\frac{\textrm{BR}(D^+\to\omega e^+\nu)}
                                       {\textrm{BR}(D^+\to\overline{K}^{*0}e^+\nu)}. $
   The $\sigma_{D^+\to\omega e\nu}$ error used in Eq.~\ref{penalty} ~is based on 
   the errors in the branching fractions reported by CLEO-c with  
   statistical and systematic errors added in quadrature to the error in 
   the efficiency corrected yield for \kstar. 
   
   $Y_{D^+_s}$ is the combined yield of the modes $\dsetaprimemunu$, $\dsetamunu$,
   and $\phimunu$ ~with $\eta'$ ~decaying to either $\rho^0\gamma$ or to $\eta\pi^+\pi^-$,
   $\eta$ ~decaying to either $\pi^+\pi^-\pi^0$ or to $\pi^+\pi^-\gamma$, and $\phi$ decaying to
   $\rho\pi$. These modes are generated simultaneously with
   their corresponding relative branching ratios, and a single shape, $S_{D^+_s}$, is obtained.

   $Y_{\mathcal{C}}$ is the number of combinatorial background events where at least one of the 
   three charged tracks forming the decay vertex does not belong to the vertex.
   After applying our selection criteria, this background is dominated by charm decays. 
   In order to generate $S_{\mathcal{C}}$, the combinatorial background shape,
   we use a large Monte Carlo sample which simulates all known charm decays, where 
   after an event is selected the reconstructed tracks are matched against the generated tracks. 
   If one of the reconstructed tracks does not belong to the generated decay vertex, the event is 
   flagged as a combinatorial background event.
        
   The last term of Eq.~\ref{mll}, $Y_{\mathcal{M}}$, is the number of events due to muon
   misidentification. The muon misidentification shape is also obtained from a large 
   Monte Carlo sample where all known charm decays are simulated. In this case tracks within 
   the acceptance of the inner muon system with a confidence level less than 1\%
   and momentum greater than 10~GeV/$c$ are taken as muons. This allows us to use the 
   same selection as in the analysis, but with very few real semi-muonic decays in the sample. 
   This shape is then weighted with a momentum-dependent misidentification probability 
   function to obtain the final shape used in the fit. The same technique, applied to a sub-sample
   of the FOCUS data, gave a shape in very good agreement with the shape used in the fit.
   We choose not to use shape or $Y_{\mathcal{M}}$ estimates from the data due to the limited statistics 
   available. Instead, we allow $Y_{\mathcal{M}}$ to float freely in the fit.

   The \kpimunu ~yield used for the normalization is estimated in the same way as 
   the \rhomunu ~yield. In this case, we only need two components in 
   the fit: one for the signal and one for the background. This background shape is 
   obtained from a Monte Carlo sample where all known charm decays, except \kpimunu, are
   generated. The fit to the $K^-\pi^+$ line shape is similar to the one used in
   \cite{vector_pseudov},
   and though this fit does not describe the complex line shape presented in \cite{kstar_spectrum},
   we find that it provides a robust estimate of the \kpimunu ~yield.
  
\begin{figure}[t!]
   \begin{center}
   \includegraphics[height=2.5in]{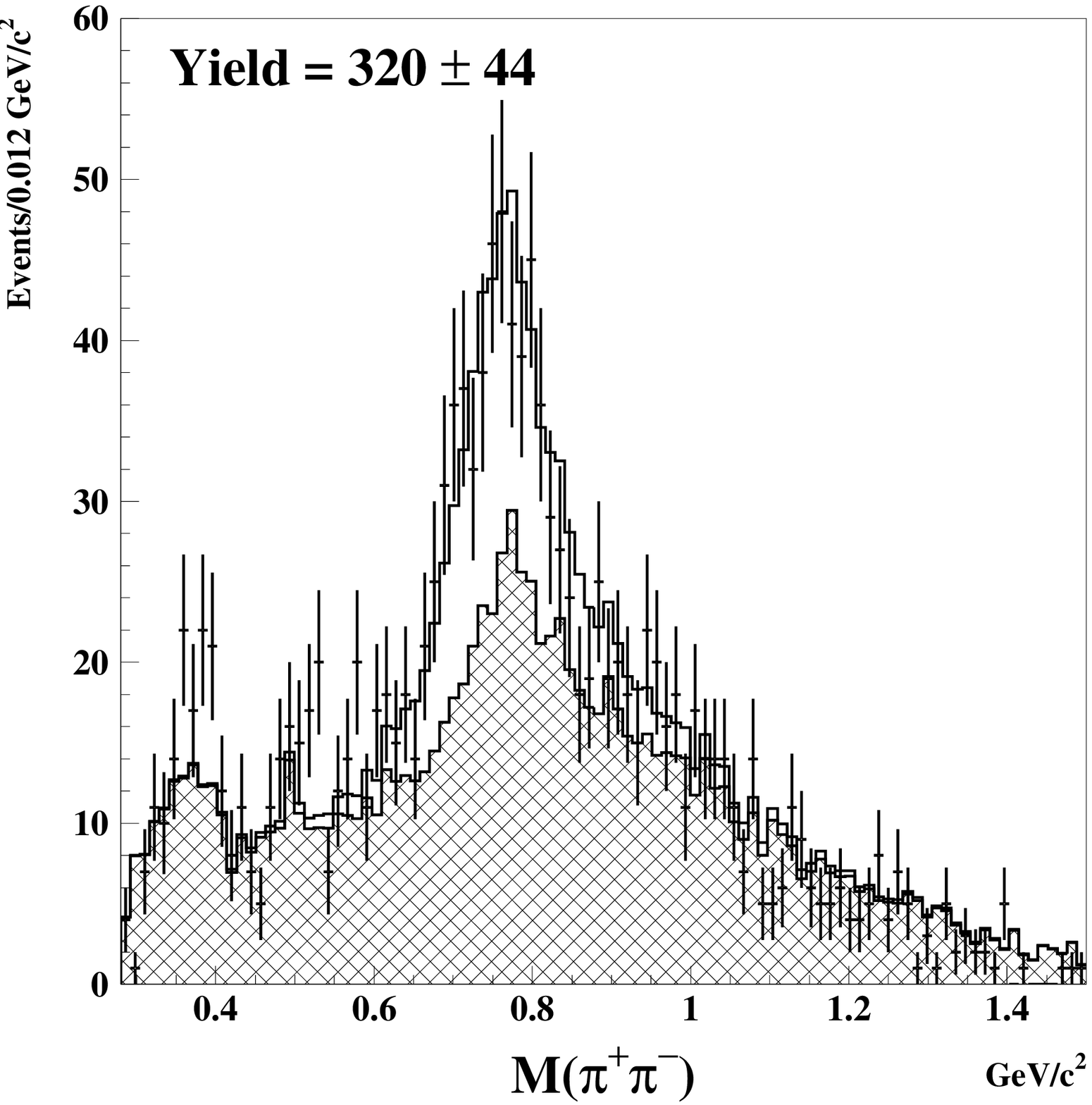}%
   \includegraphics[height=2.5in]{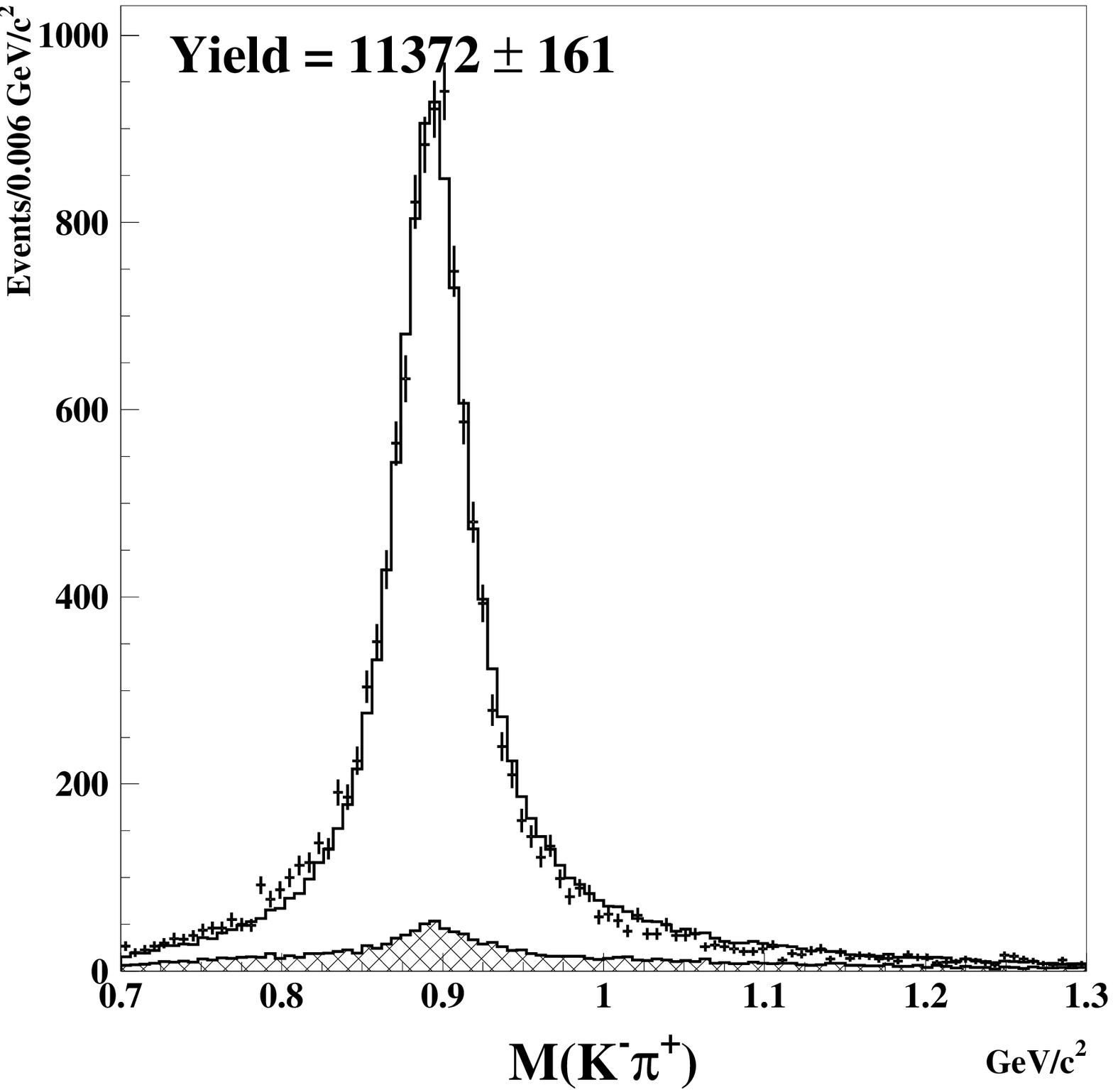}
   \caption[\rhomunu ~fit.]
   {\small{Fit to the $\pi^+\pi^-$ invariant mass (left) and the $K^-\pi^+$ invariant mass
   (right). The fit to the data (error bars) is shown as a solid line. The hatched histogram 
   on invariant mass plots is the background.}}
   \label{rho_fit}
   \end{center}
   \end{figure}

   From our fit we find $320\pm44$ \rhomunu ~events and $11,372\pm161$ $\kpimunu$ events. 
   In Fig.~\ref{rho_fit} we show the fit result. The ratio of branching ratios is 
   defined as
   \begin{equation}									 
      \frac{\textrm{BR}(\rhomunu)}{\textrm{BR}(\kstar)}=\frac{Y_{\rhomunu}/\epsilon_{\rhomunu}}		 
      {Y_{\kpimunu}/\epsilon_{\kpimunu}} \times \textrm{BR}(\overline{K}^{*0}\to K^-\pi^+).  	 
      \label{eqn:rhobr} 								 
   \end{equation}								           

   This branching ratio must be corrected to account for the  
   $(5.30\pm 0.74^{+0.99}_{-0.96})$\% non-resonant S-wave contribution present in the 
   $\kpimunu$ spectrum previously reported by FOCUS \cite{interference,kstar_spectrum}. 
   Including this correction, with errors added in quadrature, we find   
   \begin{displaymath}								           
        \frac{\textrm{BR}(\rhomunu)}{\textrm{BR}(\kstar)}=0.0412\pm0.0057.
   \end{displaymath} 								           
    
   \begin{table}[tbh!]								           
   \centering									           
   \begin{minipage}{\textwidth}
   \centering
   \begin{tabular}{|l|c|c|}\hline \hline					           
   Decay Mode		       $$      & Total Yield   	& Yield in signal region  \\ \hline				 
   $\rhomunu$			       & $320\pm44$	&  282\\  				 
   $\kpimunu$			       & $68$\footnote{The \kpimunu~yield is not a fit parameter, instead it is estimated based on
	 the efficiency for a \kpimunu ~event to be reconstructed as a \rhomunu ~event as described in
	 the text.}    	&  44\\					 
   $D^+\to K^0_S\mu^+\nu$              & $7\pm6$  	&  0\\					 
   $D^+_s$ ~ modes total	       & $179\pm40$	&  101\\					 
   $D^+\to\omega\mu^+\nu$              & $51\pm22$      &  10\\
   Muon Mis-Id       $$ 	       & $550\pm44$	&  263\\				           
   Combinatoric      $$ 	       & $233\pm50$	&  99\\ \hline \hline		           
   \end{tabular}
   \end{minipage}								           
   \caption[Contributions to \rhomunu ~signal]			           
   	 {\small{Contributions to the $\pi^+\pi^-$ invariant mass spectrum. The third column
	 shows the number of events in the signal region defined as
	 $0.62$~GeV/$c^2$\,$<$\,$M(\pi^+\pi^-)$\,$<$\,$0.92$~GeV/$c^2$.}}       
   \label{new_semilep}								           
   \end{table}									            										           
   
   Many tests were performed to ensure that the final result is stable to our detailed cut choice, as
   well as a good representation of the data. The quantitative tests we performed to
   determine additional sources of uncertainty are detailed below.
   
   \begin{figure}[t!]
   \begin{center}
   \includegraphics[height=2.5in]{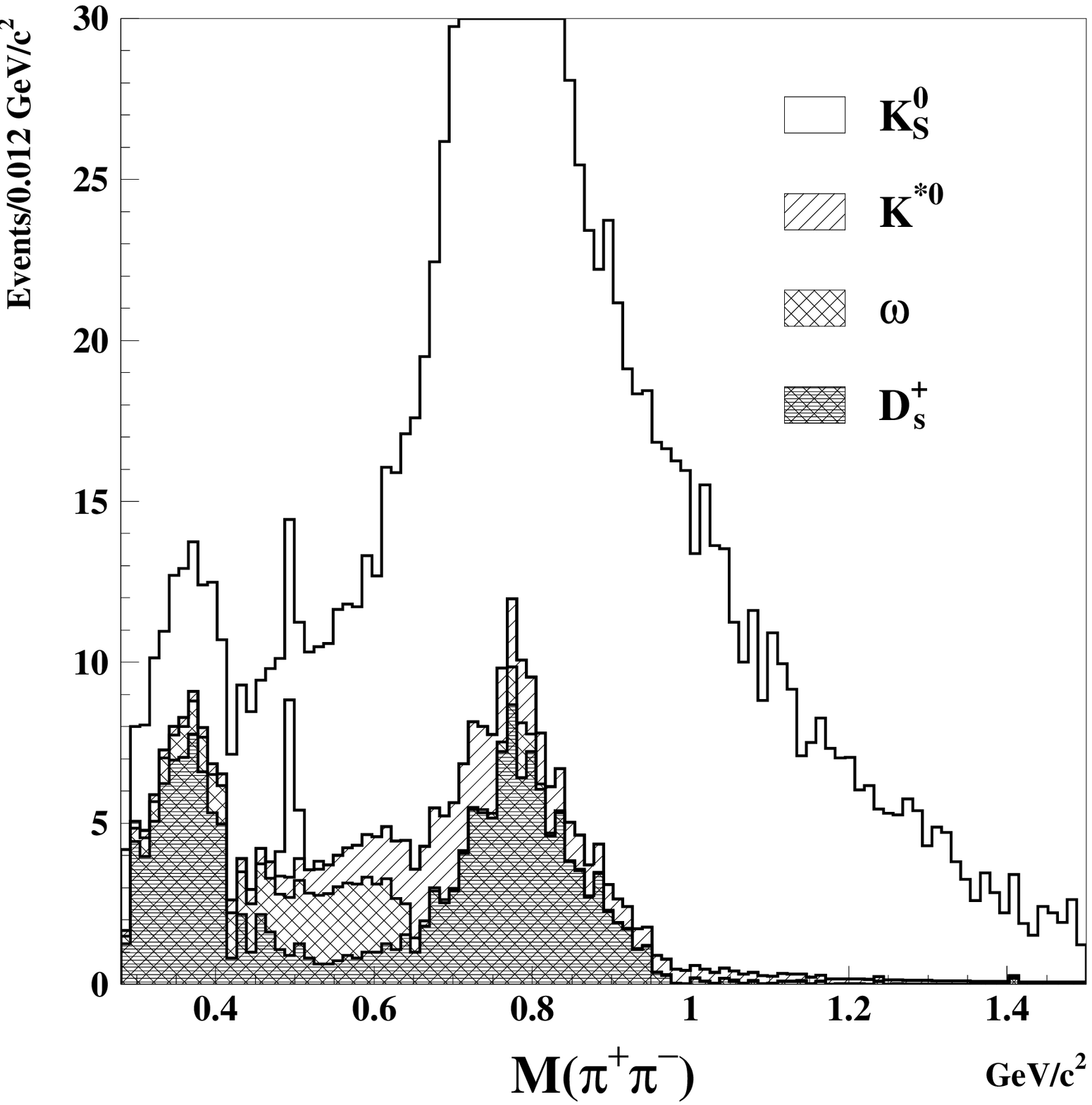}%
   \includegraphics[height=2.5in]{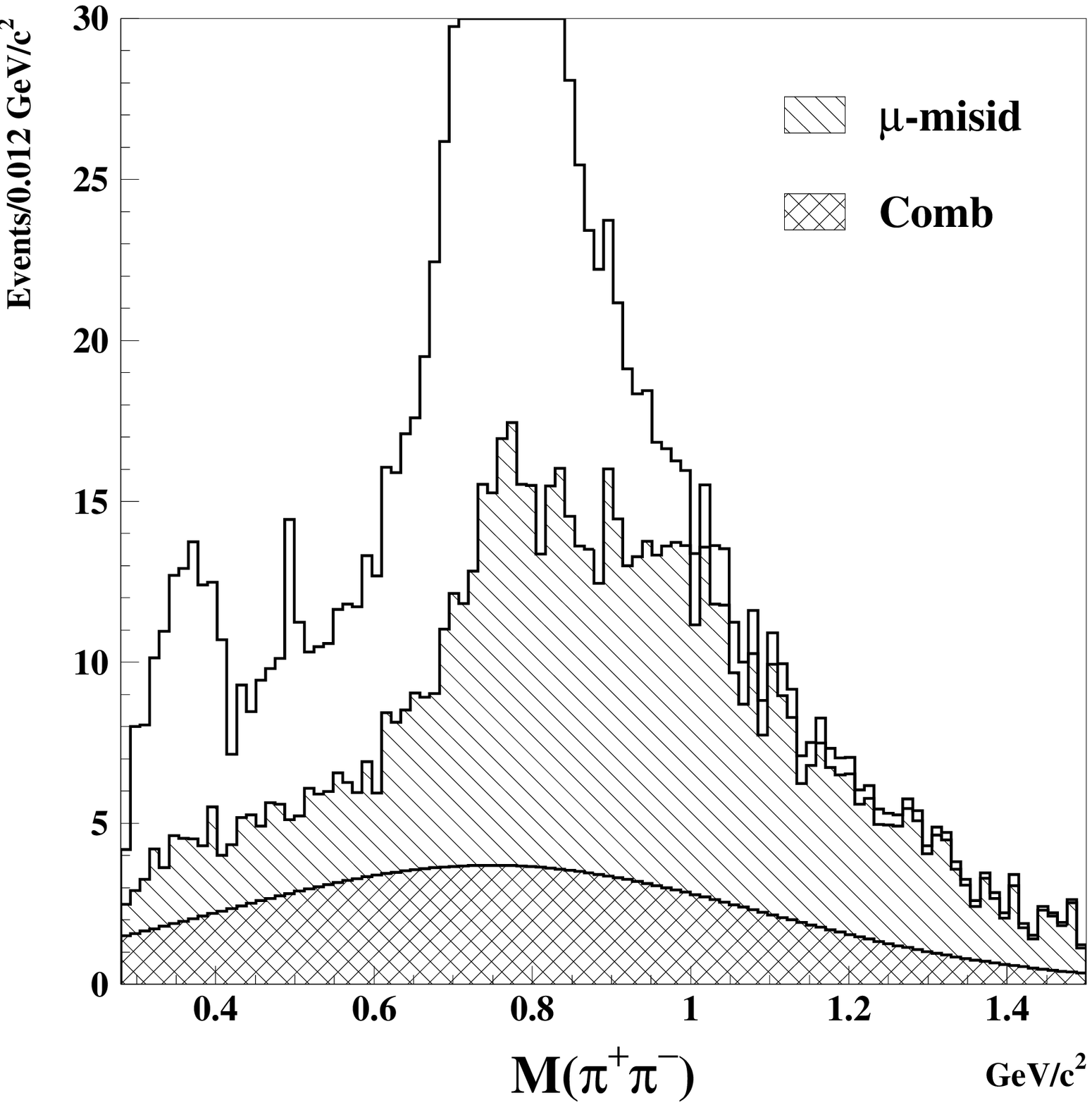}
   \caption[M($\pi^+\pi^-$) spectrum]
   {\small{M($\pi^+\pi^-$) background contributions shown in cumulative plots. 
   Left: semileptonic contributions. Right: Muon misid and combinatorial background 
   contributions. In the case of the \kstar ~and combinatorial backgrounds,
   smoothed shapes been used. The fit histogram is shown in both plots for reference.}}
   \label{semilep}
   \end{center}
   \end{figure}

\section{Systematic Studies}
   
   Several studies have been performed in order to assess systematic
   contributions to the uncertainty in the ratio. Three possible contributions have been 
   identified.
   
   The first contribution is due to the final cut selection  used to determine the branching ratio. 
   In order to estimate a contribution from this source, we vary our cuts around the final choice to
   exercise likely differences between the signal and background.  Since the $D^+$ is longed lived
   compared to sources of background from $D^+_s$, and other short-lived backgrounds such as those
   coming from non-charm sources, we vary the significance of the separation between the production 
   and the decay vertices from $10\sigma$ to $20\sigma$, and out of target requirements for the decay vertex
   from $0\sigma$ to $2\sigma$. To look for poorly
   formed vertices and vertices that are formed from particles that decay into muons early in the
   spectrometer, we vary  the confidence level of the secondary vertex from 1\% to 10\%.  We have
   estimated the feed-down from $D^+\to K^-\pi^+\mu^+\nu$ using our Monte Carlo simulation, but look for
   backgrounds we might have missed by varying the \cer
   ~identification cuts for the pions from 4 to 6 units of likelihood. The level of the muon
   misid was checked by changing the muon identification confidence level from 1\% to 10\%, the muon
   momentum cut from 10~GeV/$c$ to 20~GeV/$c$, and selecting events that left hits in all 6 of the muon planes.
   A very stringent test which dramatically changes the background level is to relax the visible mass
   cut. Though the statistical significance of the result suffers due to the inclusion of so much
   background, this is an important check on backgrounds we might have missed coming from higher
   multiplicity modes, which are expected to be small, and combinatorial sources.
   
   The cut systematic is assessed by
   measuring the branching ratio for the different cut combinations and calculating the sample
   variance for the returned values of \rhobr. Because our tested cuts have succeeded in delivering a
   broad range of signal to background values as well as changes in the final yield, this method is
   likely to deliver a conservative estimate of the systematic error due to our cut selection. We find
   no significant change in the branching ratio due to our particular cut choice and assign a 
   systematic uncertainty of 0.0023 due to our cut selection.
   
   The second contribution is related to our fit. In order to check the fit
   against possible biases as well as the accuracy of the statistical error reported by the
   fit, we perform the fit multiple times after fluctuating each bin of the data
   histogram using a Poisson distribution. We find that both the mean and width of the
   distribution of the fit results are in agreement with our reported results.
   
   We test the effect of the fit inputs by changing the \kpimunu ~ECY used
   to estimate the amount of background due to $K/\pi$ misidentification in our \rhomunu 
   ~sample by a factor of two and by fitting with no restrictions on the
   $D^+\to\omega\mu^+\nu$ yield. We test the combined $D^+_s$ shape used in the
   fit by fitting our signal using the individual shapes of the $D^+_s$ modes mentioned 
   earlier. In this case, we replace the $D^+_s$ yield parameter in the fit with a parameter
   representing the \phimunu ~ECY and we extract the individual yields using the branching
   ratios of these modes relative to \phimunu. These branching ratios are then varied
   by $\pm 1\sigma$. We have also changed the shape of the combinatorial background by
   replacing it with the shape obtained when two same sign pions are used to form a $\pi\pi\mu$
   vertex. As a final check on the fit, we have changed the binning scheme and mass range 
   used in the fit. As in the case of cut variations, we calculate the sample variance of the 
   returned values and quote this as our systematic contribution. We find this contribution 
   to be 0.0038 mostly coming from the uncertainty in the combinatorial background shape.
   
   The third category includes a search for additional, unaccounted for, systematic uncertainty that may
   come from the detector simulation and/or the charm production mechanism. We estimate this  by splitting
   our sample into three pairs of statistically independent sub-samples. A powerful test for the production 
   model, trigger, and detector simulation for the signal as well as for the background is to split the 
   data according to the $D^\pm$ momentum. Another test of the production model is to look whether we 
   have a decay of a $D^+$ or a $D^-$. Our final split sample looks at two different detector 
   configurations. For roughly 30\% of the FOCUS running we ran without the interleaved silicon planes 
   within the target. With this split, we tested the detector simulation as well as lifetime dependent
   backgrounds. 
   
   In order to separate the likely systematic error contribution in any
   differences in the results from larger statistical fluctuations due to reduced statistics,
   FOCUS uses a technique based on the {\it S-factor} method of the PDG. In this method the 
   branching ratio is measured for each pair of split sub-samples and a scaled variance is calculated. 
   The split sample contribution is the difference between the scaled variance and the
   statistical variance if the scaled variance is greater than the statistical
   variance for the entire sample. We find that no additional contribution to the systematic
   uncertainty is indicated by this search.
   
   The total systematic uncertainty is obtained by adding in quadrature all of these
   contributions as summarized in Table~\ref{syst}.
   \begin{table}[h]
   \centering
   \begin{tabular}{||l|c||}	           \hline \hline
      Systematic Source &   Error 	\\ \hline
      Cut variations	&  0.0023        \\
      Fit variation	&  0.0038        \\
      Total		&  0.0044 	\\ \hline \hline
   \end{tabular}
   \caption[Systematic error]{\small{Sources of systematic errors. The three sources
	   are added in quadrature to obtain the total systematic error.}}
   \label{syst}	   
   \end{table}

\section{Conclusions}      
    From $320\pm 44$ \rhomunu ~decays and $11,372\pm 161$ \kpimunu ~decays, we report a
    measurement of the branching ratio
       \begin{displaymath}								           
           \frac{\textrm{BR}(\rhomunu)}{\textrm{BR}(\kstar)}=%
	         0.041\pm0.006~\textrm{(stat.)}\pm0.004~\textrm{(syst.)}.
        \end{displaymath} 								           
    Using this result along with the FOCUS measurement of the ratio \kstarbr~\cite{kstar_focus}, 
    the PDG~\cite{pdg} value of the absolute branching fraction of the decay $\kpipi$, and the FOCUS 
    measurement of the $D^+$ lifetime~\cite{focus_lifetime}, we calculate     
    \begin{displaymath}								           
        \Gamma(\rhomunu)=(0.22\pm0.03\pm0.02\pm0.01)\times 10^{10}~\textrm{s}^{-1}
    \end{displaymath} 								           
    where the last error is a combination of the uncertainties on the quantities not measured in
    this work. When calculating the partial decay width, we have corrected \kstarbr ~with the
    updated value for the S-wave non-resonant contribution~\cite{kstar_spectrum}.
    In Table~\ref{exp} and Table~\ref{predictions}, we compare our result to previous
    experimental results and theoretical predictions, respectively.
    
    Our result for \rhobr is consistent with a recent CLEO collaboration result on the absolute 
    branching ratios for $D^+$ semi-electronic decays~\cite{cleo_d+} and represents a significant
    improvement to the world average for the semi-muonic mode. The experimental results indicate 
    that the QCD Sum  Rule predictions for \rholnu~\cite{ball,yang_hwang} are too low.

   \begin{table}[htb!]
   \centering
   \begin{tabular}{||l|c|c||} \hline \hline
   Reference			& $\frac{\textrm{BR}(\rhomunu)}{\textrm{BR}(\kstar)}$ %
                                & $\frac{\textrm{BR}(\rhoenu)}{\textrm{BR}(\kstarenu)}$\\  \hline
   E653~\cite{e653_rho}         & $0.044^{+0.034}_{-0.029}$  &       \\
   E687~\cite{e687_rho} 	& $0.079\pm0.023$          &       \\
   E791~\cite{e791_rho}         & $0.051\pm0.017$          &  $0.045\pm0.017$     \\
   CLEO~\cite{cleo_d+}		&        		     &  $0.038\pm0.008$       \\
   This result                  & $0.041\pm0.007$          &       \\ \hline \hline
   \end{tabular}
   \caption[Experimental Results]
   	 {\small{Experimental results for the branching ratio. Statistical and systematic
	 errors have been added in quadrature.}}
   \label{exp}
   \end{table}

    \begin{table}[htb!]
    \centering
    \begin{tabular}{||l|c|c|c||}		     						                \hline \hline
    	Reference 	  			    & $\ell$ &
	$\frac{\textrm{BR}(\rholnu)}{\textrm{BR}(\kstarlnu)}$& $\Gamma(\rholnu) (10^{10}~s^{-1})$\\ \hline
	Ball~\cite{ball}  (SR)                       & $e$  &				     & $0.06\pm0.02$ \\
	APE~\cite{ape}	  (LQCD)    	             & $\ell$  & $0.043\pm0.018$	     & $0.3\pm0.1$\\   
    	Jaus~\cite{jaus}  (QM)		             & $\ell$  & 0.030  		     & 0.16 \\ 
	ISGW2~\cite{isgw2} (QM)		             & $e$  & 0.023			     & 0.12 \\
    	Yang--Hwang~\cite{yang_hwang}	(SR)         & $e$  & $0.018\pm0.005$		     & $0.07^{+0.04}_{-0.02}$ \\ 
	O'Donnell--Turan~\cite{odonnell_turan} (LF)  & $\mu$  & 0.025			     &  \\		    
    	Melikhov~\cite{melikhov}  (QM)               & $\ell$  & 0.027, 0.024		     & 0.15, 0.13 \\  
	Ligeti--Stewart--Wise~\cite{ligeti}          & $\ell$  & 0.044                       & \\
        Kondratyuk-Tchein~\cite{kondratyuk}  (LF)    & $\ell$  & 0.035, 0.033, 0.033, 0.032  & 0.19, 0.20, 0.18, 0.19 \\
	Melikhov--Stech~\cite{melikhov-stech}  (QM)  & $\ell$  & 0.035  		     &0.21 \\ 
	Wang--Wu--Zhong~\cite{wang_wu_zhong} (LC)    & $\ell$  & $0.035\pm0.011$	     &$0.17\pm0.04$\\	
	Fajfer--Kamenik~\cite{kamenik} 		     & $\ell$  & $0.045$ 	     &$0.25$\\	
    	This result                                  & $\mu$  & $0.041\pm0.007$ 	     & $0.22\pm0.04$ \\\hline \hline 
    \end{tabular}
   \caption[Theoretical predictions]
   	 {\small{Theoretical predictions for the branching ratio
	         and partial decay width. Most of the
		 theoretical predictions are calculated for $D^0\to\rho^-\ell^+\nu$. To 
		 compare these predictions with our result, we have used the
		 isospin conjugate relation 
		 $\Gamma(D^+\to\rho^0\ell^+\nu)=1/2~\Gamma(D^0\to\rho^-\ell^+\nu)$.}}
   \label{predictions}
   \end{table}
   
\section{Acknowledgments}

   We wish to acknowledge the assistance of the staffs of Fermi National
   Accelerator Laboratory, the INFN of Italy, and the physics departments
   of the collaborating institutions. This research was supported in part
   by the U.~S.  National Science Foundation, the U.~S. Department of
   Energy, the Italian Istituto Nazionale di Fisica Nucleare and
   Ministero dell'Universit\`a e della Ricerca Scientifica e Tecnologica,
   the Brazilian Conselho Nacional de Desenvolvimento Cient\'{\i}fico e
   Tecnol\'ogico, CONACyT-M\'exico, the Korean Ministry of Education, and
   the Korea Research Foundation.

\bibliographystyle{elsart-num}

\begin{thebibliography}{10}
\expandafter\ifx\csname url\endcsname\relax
  \def\url#1{\texttt{#1}}\fi
\expandafter\ifx\csname urlprefix\endcsname\relax\def\urlprefix{URL }\fi

\bibitem{ball}
{P. Ball}, {Semileptonic Decays
  $D\to\pi(\rho)e\nu$~and~$B\to\pi(\rho)e\nu$~from QCD Sum Rules}, {Phys. Rev.
  D} 48 (1993) 3190.

\bibitem{ape}
{C. R. Allton {et al.}}, {Lattice Calculation of D~and~B~Meson Semileptonic
  Decays, Using the Clover Action at $\beta~=~6.0$~on APE}, {Phys. Lett. B} 345
  (1995) 513.

\bibitem{jaus}
{W. Jaus}, {Semileptonic, Radiative and Pionic decays of B, B* and D, D*
  Mesons}, {Phys. Rev. D} 53 (1996) 1349.

\bibitem{isgw2}
{N. Igsur and D. Scora}, {Semileptonic Meson Decays in the Quark Model: An
  Update}, {Phys. Rev. D} 52 (1995) 2783.

\bibitem{yang_hwang}
{W. Y. Wang, Y. L. Wu and M. Zhong}, {The QCD Sum Rule Approach for the
  Semileptonic Decay of the D or B Meson into a light meson and leptons}, {Z.
  Phys, C} 73 (1997) 275.

\bibitem{odonnell_turan}
{P. J. O'Donnell and G. Turan}, {Charm and Bottom Semileptonic Decays}, {Phys.
  Rev. D} 56 (1997) 295.

\bibitem{melikhov}
{D. Melikhov}, {Exclusive Semileptonic Decays of Heavy Mesons in the Quark
  Model}, {Phys. Lett. B} 394 (1997) 385.

\bibitem{ligeti}
{Z. Ligeti, I. W. Stewart and M. B. Wise}, {Comment on $V_{ub}$ from Exclusive
  Semileptonic $B$ and $D$ Decays}, {Phys. Lett. B} 420 (1998) 359.

\bibitem{kondratyuk}
{L. A. Kondratyuk and D. V. Tchekin}, {Transition Form Factors and
  Probabilities of the Semileptonic Decays of B and D Mesons within Covariant
  Light-Front Dynamics}, {Phys. Atom. Nucl.} 64 (2001) 727.

\bibitem{melikhov-stech}
{D. Melikhov and B. Stech}, {Weak Form Factors for Heavy Meson Decays: An
  Update}, {Phys. Rev. D} 62 (2000) 014006.

\bibitem{wang_wu_zhong}
{W. Y. Wang, Y. L. Wu and M. Zhong}, {Heavy to Light Meson Exclusive
  Semileptonic Decays in Effective Field Theory of Heavy Quarks}, {Phys. Rev.
  D} 67 (2003) 014024.

\bibitem{kamenik}
{S. Fajfer and J. Kamenik}, {Charm Meson Resonances and $D\to V$ Semileptonic
  Form Factors}, {Phys. Rev. D} 72 (2005) 034029.

\bibitem{e653_rho}
{K. Kodama {et al.}}, {Observation of $D^+\to\rho(770)^0\mu^+\nu$}, {Phys.
  Lett. B} 316 (1993) 455.

\bibitem{e687_rho}
{P. L. Frabetti {et al.}}, {Observation of the Vector Meson Cabibbo Suppressed
  Decay \rhomunu}, {Phys. Lett. B} 391 (1997) 235.

\bibitem{e791_rho}
{E. M. Aitala {et al.}}, {Measurement of the Branching Ratio
  $\frac{\textrm{B}(D^{+}\to \rho^{0}l^{+}\nu_l)} {\textrm{B}(D^{+}\to
  \overline{K^{*0}}l^{+}\nu_l)}$}, {Phys. Lett. B} 397 (1997) 325.

\bibitem{pdg}
{S. Eidelman {et al.}}, {Review of Particle Physics}, {Phys Lett. B} 592 (2004)
  1.

\bibitem{e687_det}
{P. L. Frabetti {et al.}}, {Description and Performance of the Fermilab E687
  Spectrometer}, {Nucl. Instrum. Meth. A} 320 (1992) 519.

\bibitem{TS}
{J. M. Link {et al.}}, {The Target Silicon Detector for the FOCUS
  Spectrometer}, {Nucl. Instrum. Meth. A} 516 (2004) 364.

\bibitem{Cer}
{J. M. Link {et al.}}, {\cer ~Identification in FOCUS}, {Nucl. Instrum. Meth.
  A} 484 (2002) 270.

\bibitem{interference}
{J. M. Link {et al.}}, {Evidence for New Interference Phenomena in the Decay
  $D^+ \to K^-\pi^+\mu^+\nu$}, {Phys. Lett. B} 535 (2002) 43.

\bibitem{pythia}
{T. Sj\"{o}strand}, {High Energy Physics Event Generation with PYTHIA 5.7 and
  JETSET 7.4}, {Comp. Phys. Comm} 82 (1994) 74.

\bibitem{cleo_d+}
{G. S. Huang {et al.}}, {Absolute Branching Fraction Measurements of Exclusive
  $D^+$ Semileptonic Decays}, {hep-ex/0506053} .

\bibitem{vector_pseudov}
{J. M. Link {et al.}}, {Measurement of the Ratio of the Vector to Pseudoscalar
  Charm Semileptonic Decay Rate $\frac{\Gamma(D^+ \to \overline{K}^{*0} \mu^+
  \nu)}{\Gamma(D^+ \to \overline{K}^0 \mu^+ \nu)}$}, {Phys. Lett. B} 598 (2004)
  33.

\bibitem{kstar_spectrum}
{J. M. Link {et al.}}, {Hadronic Mass Spectrum Analysis of $D^+\to
  K^-\pi^+\mu^+\nu$ Decay and Measurement of the $K^*(892)^0$ Mass and Width},
  {Phys. Lett. B} 320 (2005) 72.

\bibitem{kstar_focus}
{J. M. Link {et al.}}, {New Measurement of the
  $\frac{\Gamma(D^+\to\overline{K}^{*0}\mu^+\nu)} {\Gamma(D^+\to
  K^-\pi^+\pi^+)}$ and $\frac{\Gamma(D^+_S \to \phi \mu^+ \nu)}{\Gamma(D^+_S
  \to \phi \pi^+)}$}, {Phys. Lett. B} 541 (2002) 243.

\bibitem{focus_lifetime}
{J. M. Link {et al.}}, {New Measurement of the $D^0$ ~and $D^+$ lifetimes},
  {Phys. Lett. B} 537 (2002) 192.

\end{thebibliography}

\end{document}